# Cooling flows, central galaxy–cluster alignments, X-ray absorption and dust

S.W. Allen[1], A.C. Fabian[1], A.C. Edge[1], H. Böhringer[2], D.A. White[1]

[1]*Institute of Astronomy, Madingley Road, Cambridge CB3 OHA*
[2]*Max-Planck-Institut für extraterrestrische Physik, D85740 Garching b. München, Germany*

**ABSTRACT**
We present the analysis of pointed ROSAT PSPC observations of five of the most luminous, intermediate redshift ($0.1 < z < 0.15$) clusters of galaxies detected in the ROSAT All-Sky Survey. The PSPC data are combined with optical CCD images and spectra to examine the relationship between clusters and their central cluster galaxies (CCGs). Abell 1068, Abell 1361 and Abell 1664 contain three of the most optically line-luminous CCGs known. The PSPC X-ray data show that these galaxies lie at the centres of massive ($200-400$ $M_\odot$ yr$^{-1}$ ) cooling flows. The alignment between CCGs and their host clusters has been investigated. For those clusters with cooling flows, the position angles of the X-ray emission from the clusters and the optical emission from the CCGs agrees within 5 degrees. For the one probable non-cooling flow cluster in the sample, Abell 2208, the alignment is significantly poorer. We examine the evidence for intrinsic X-ray absorption in the clusters. The X-ray spectra for Abell 1068 and Abell 1664 show that the cooling flows in these clusters are intrinsically absorbed by equivalent hydrogen column densities $\gtrsim 10^{21}$ atom cm$^{-2}$. The optical spectra of the CCGs in these clusters exhibit substantial intrinsic reddening, at levels consistent with the X-ray absorption results if standard dust to gas ratios are assumed.

## 1 INTRODUCTION

The ROSAT X-ray All-Sky Survey (RASS; Voges 1992), which took place between 1990 Aug and 1991 Jan, has identified more than fifty thousand new X-ray sources. One of the primary aims of the RASS is a study of the distribution and evolution of clusters of galaxies. In earlier papers (Allen *et al.* 1992; Crawford *et al.* 1995; see also Ebeling *et al.* 1993) we have described the X-ray selection of, and optical follow-up study to, the ROSAT Brightest Cluster Sample (BCS). The BCS is the flux-limited catalogue of the $\sim 300$ X-ray brightest clusters at declinations $> 0$ deg, Galactic latitude, $b > +20$ deg, and RASS counts rates $\gtrsim 0.2$ ct s$^{-1}$. An integral part of the BCS study is an examination of the interrelation between the X-ray properties of clusters and the optical properties of their central galaxies. Central cluster galaxies, the optically-dominant galaxies in the cores of clusters (referred to throughout this paper as CCGs), frequently exhibit unusual optical properties. In particular, earlier studies (Edge, Stewart & Fabian 199; Donahue *et al.* 1992; Allen *et al.* 1992; Crawford *et al.* 1995) have shown that $30-40$ per cent of CCGs exhibit optical line-emission which, in nearby systems, is seen to form extended filamentary nebulae (with radial extents of $\sim 10$ kpc; *e.g.* Johnstone & Fabian 1988; Heckman *et al.* 1989; Crawford & Fabian 1992). The luminosity in H$\alpha$ emission alone exceeds $10^{42}$ erg s$^{-1}$ in the most extreme cases (Allen *et al.*

1992). The optical properties of CCGs are intimately related to the properties of the surrounding Intracluster Medium (hereafter ICM: Heckman *et al.* 1989; Crawford & Fabian 1992). Optical line emission is only observed in the cores of clusters with cooling flows, although not all clusters for which the X-ray data indicate the presence of a cooling flow exhibit optical line emission.

This paper describes the X-ray and optical properties of a subsample of BCS clusters in the redshift range $0.1 < z < 0.15$. We have used the Position Sensitive Proportional Counter (PSPC) on ROSAT to examine the evidence for cooling flows in the clusters, and to investigate intrinsic X-ray absorption within the flows (White *et al.* 1991; Allen *et al.* 1993). The clusters discussed in most detail; Abell 1068, Abell 1664, Abell 1361 and Abell 2208 are all Abell richness class I, and have optical morphologies ranging from Bautz-Morgan (BM) type I for Abell 1068, types I–II for Abell 1361 and Abell 2208, through to type II–III for Abell 1664. [The X-ray image of a fifth cluster, Abell 1413 (also in the BCS; Abell richness class 3, BM type I, and with a redshift $z = 0.1427$) was extracted from the ROSAT Data Archive and is included in the sample for comparison purposes.] The clusters were selected for study on the basis of both their high RASS count rates and for the wide range of optical properties exhibited by their CCGs; Abell 1068 and Abell 1664 contain the two most optically line-luminous CCGs in the $0.1 < z < 0.15$ interval of the BCS (both galaxies have H$\alpha$ luminosities $\gtrsim 10^{42}$ erg s$^{-1}$ ). The line ratios observed in the emission-line gas surrounding the two galaxies indicate quite different states of ionization, however. For Abell 1664, H$\alpha$ > [NII] and [OII] > [OIII], whereas Abell 1068 has [NII] > H$\alpha$ and [OIII] > [OII] (further details of the optical CCG spectra are presented by Allen *et al.* 1992.) The CCG of Abell 1361 also exhibits emission lines but with a H$\alpha$ luminosity an order of magnitude less than Abell 1068 or Abell 1664. No optical emission lines are observed in the CCGs of Abell 2208 and Abell 1413. (Details of the optical spectrum of Abell 1413 were kindly provided by A. Dunn.)

CCGs normally have recession velocities very near the mean values for their host clusters and are thought to rest at the base of the cluster potentials. [For discussion of these points see Oegerle & Hill 1994 and references therein.] Optical observations of CCGs show that in general they are more elongated than other elliptical galaxies and have their axes aligned with those of their host clusters (Carter & Metcalfe 1980; Dressler 1981; Porter *et al.* 1991). The close relationship between the optical properties of CCGs and their host clusters, together with studies of substructure and 'multiple nuclei' in these galaxies (Lauer 1988), are taken as strong evidence that merging plays an important role in CCG formation. This paper discusses the morphological correspondence between the X-ray emission from clusters of galaxies and the optical emission from their CCGs, with comparison to the results from previous optically-based studies.

The structure of this paper is as follows: Section 2 describes the observations and data reduction. In Section 3 we present the results of the morphology analysis. Section 4 describes the results from deprojection analyses of the X-ray images and the evidence for cooling flows in the clusters. Section 5 describes a spectral analysis of the X-ray data. In Section 6 we discuss the implications of our results. A value for the Hubble constant of $H_0$=50 km s$^{-1}$ Mpc$^{-1}$ and a cosmological deceleration parameter $q_0$=0 are assumed throughout.

## 2   OBSERVATIONS

The X-ray observations were made using the Position Sensitive Proportional Counter (PSPC) on ROSAT. This instrument provides a spatial resolution of $\sim 25$ arcsec [full width at half maximum (FWHM) corresponding to a spatial scale of $60 - 90$ kpc for objects in the redshift range $z = 0.1 - 0.15$] and a (FWHM) spectral resolution of $\Delta E/E = 0.43(E/0.93\text{keV})^{-0.5}$. The details of the PSPC observations are summarized in Table 1 where we list the observation dates, raw exposure times, and the exposures after correction for satellite dead time and the removal of periods of high particle background and scattered solar X-ray contamination. X-ray contour maps (in the $0.5 - 2.4$ keV band) for the clusters are presented in Figs. 1(a)–5(a). In all cases the X-ray emission from the clusters is completely contained within the central $\sim 0.3$ deg radius circular aperture defined by the PSPC rib support structure. Reduction of the X-ray data was carried out using the STARLINK ASTERIX data reduction package.

Optical R-band images of the CCGs of Abell 1068, Abell 1361 and Abell 1413 were obtained from observations with the 1.0 m Jacobus Kapteyn Telescope (JKT), La Palma, in 1992 Jan. Images of Abell 1664 and Abell 2208 were obtained with the 2.5 m Isaac Newton Telescope, La Palma, in 1994 Jun. The fives CCG images are presented in Figs. 1(b)–5(b). Reduction and analysis of the optical data were carried out using IRAF.

## 3   MORPHOLOGY ANALYSIS

We have measured the variation of the ellipticity and position angle of the X-ray emission from the clusters with radius using the ELLIPSE isophote-fitting routines in IRAF. Models of the X-ray emission were created from least-squares fits of elliptical isophotes to the unsmoothed X-ray images (Jedrzejewski 1987) with the ellipticities, position angles and centroids as free parameters in the fits. The results are summarised in Table 2 where we list the co-ordinates of the X-ray peak and the mean ellipticities and position angles between semi-major axes of $0 - 0.5$ Mpc and $0.5 - 1$ Mpc. We find no evidence for significant variations of the X-ray centroid with radius in any of the clusters except Abell 1664, where a highly-disturbed morphology is observed at large ($\gtrsim 0.5$ Mpc) radii. The same analysis technique was applied to the optical R-band images of the CCGs. The (J2000) co-ordinates, and mean ellipticities and position angles (PAs) between semi-major axes of 2 and 10 arcsec, are listed in Table 3.

The results presented in Tables 2 and 3 show that the mean PA of the optical isophotes and the mean X-ray PA within 0.5 Mpc for Abell 1068, Abell 1361, Abell 1413 and Abell 1664 all agree to within 5 degree. The cluster and CCG ellipticities are similar, although the CCG isophotes are typically slightly more elliptical than those of the clusters.

The optical coordinates for the CCGs of Abell 1068, Abell 1361, Abell 1413 and Abell 1664 are in excellent agreement with the positions of the peaks of the cluster X-ray emission. (The error in the aspect solution of the X-ray data being $\sim 5$ arcsec for a typical observation.) For Abell 2208 the discrepancy is slightly larger ($\sim 1$ arcmin) although the X-ray emission for this cluster is substantially less sharply peaked than for the other clusters and the X-ray centroid is therefore less well defined. The agreement between the CCG coordinates and the X-ray peak coordinates is significantly better than between the CCG co-ordinates and the Abell, Corwin & Olowin (1989) cluster centres.

## 4   DEPROJECTION OF THE X-RAY IMAGES

We have analysed the X-ray images using the 'deprojection' technique described by Fabian *et al.* (1981). The free parameters in the deprojection analysis are the forms of the gravitational potentials of the clusters and the pressures of the X-ray emitting gas at a selected outer radii. The cluster potentials have been modelled by King laws, with velocity dispersions, $\sigma_r$, and core radii, $r_c$. Since no useful galaxy velocity data exist for the clusters, velocity dispersions have been adjusted to ensure that the ambient gas temperatures derived from the deprojection analysis are consistent with the spectral results (Section 5). For Abell 1361 no useful spectral constraint on the ambient temperature is available and for this cluster the velocity dispersion has been approximated using the $L_X/\sigma_r$ relation of Edge & Stewart (1991). Core radii are essentially unconstrained by the data and have been fixed at a value of 200 kpc. The pressure at the outermost edge of each cluster is chosen to produce an approximately isothermal temperature profile outside of the central cooling-flow region (at a temperature in agreement with the spectral results).

The deprojection analyses were carried out on background-subtracted, vignetting-corrected images constructed on a 30 arcsec pixel scale (approximately the spatial resolution of the PSPC) using counts in the energy range $0.5 - 2.4$ keV (*i.e.* PHA channels 50–240). In each case a uniform absorbing column density consistent

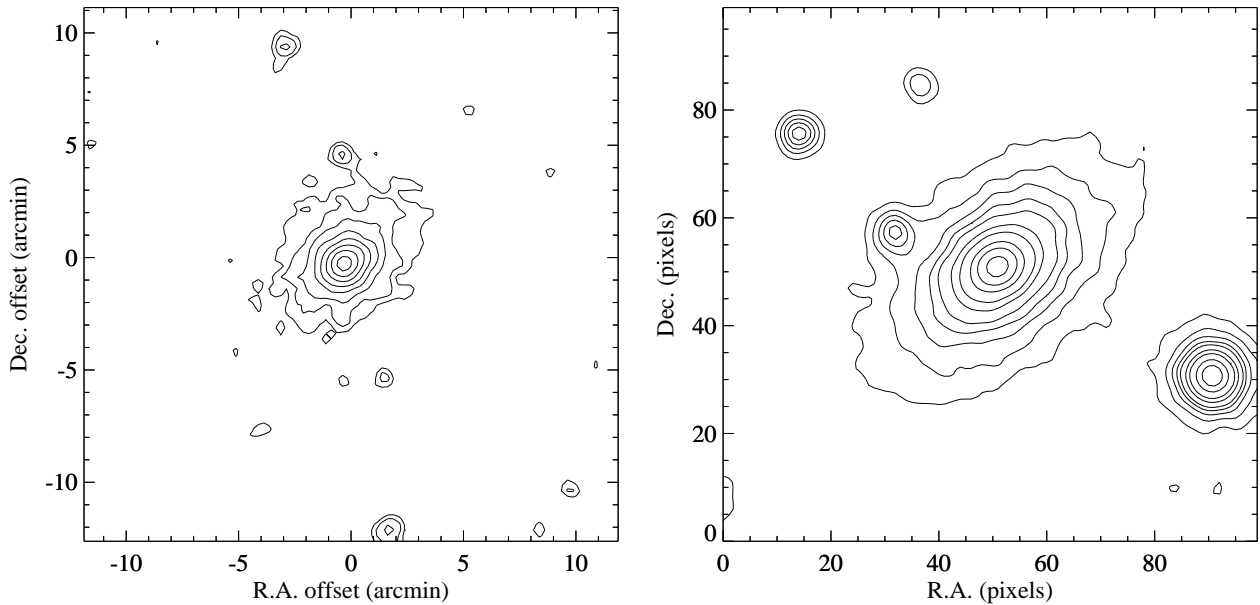

**Figure 1.** Contours of constant X-ray surface brightness for Abell 1068 in the $0.5 - 2.4$ keV band. A pixel size of $15 \times 15$ arcsec$^2$ has been used and the images have been smoothed with a gaussian of FWHM 2 pixels. The lowest contour is drawn at $10^{-4}$ count s$^{-1}$ pixel$^{-1}$ with each interior contour representing a factor two increase in X-ray brightness. For reference, at the redshift of Abell 1068 0.1 deg corresponds to 1.20 Mpc. (b) Contours of constant optical surface brightness for the CCG of Abell 1068. The optical image was obtained with the CCD camera on the 1.1 m Jacobus Kapteyn Telescope (JKT), La Palma in 1992 Jan. and has been smoothed with a gaussian of FWHM 2 pixels. The spatial scale is 0.31 arcsec pixel$^{-1}$. Note that 10 pixels $\sim 3$ arcsec which corresponds to $\sim 10$ kpc at the redshift of Abell 1068.

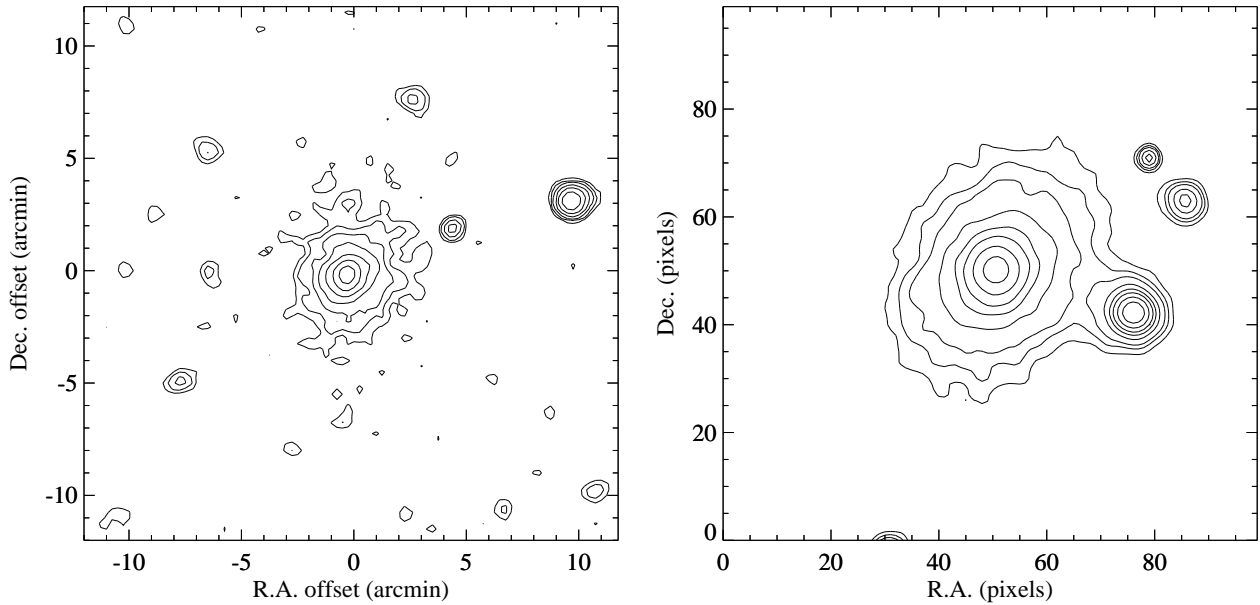

**Figure 2.** a) Contours of constant X-ray surface brightness for Abell 1361. Details as in Fig. 1a. (b) Contours of constant optical surface brightness for the CCG of Abell 1361. Details as in Fig. 1b.

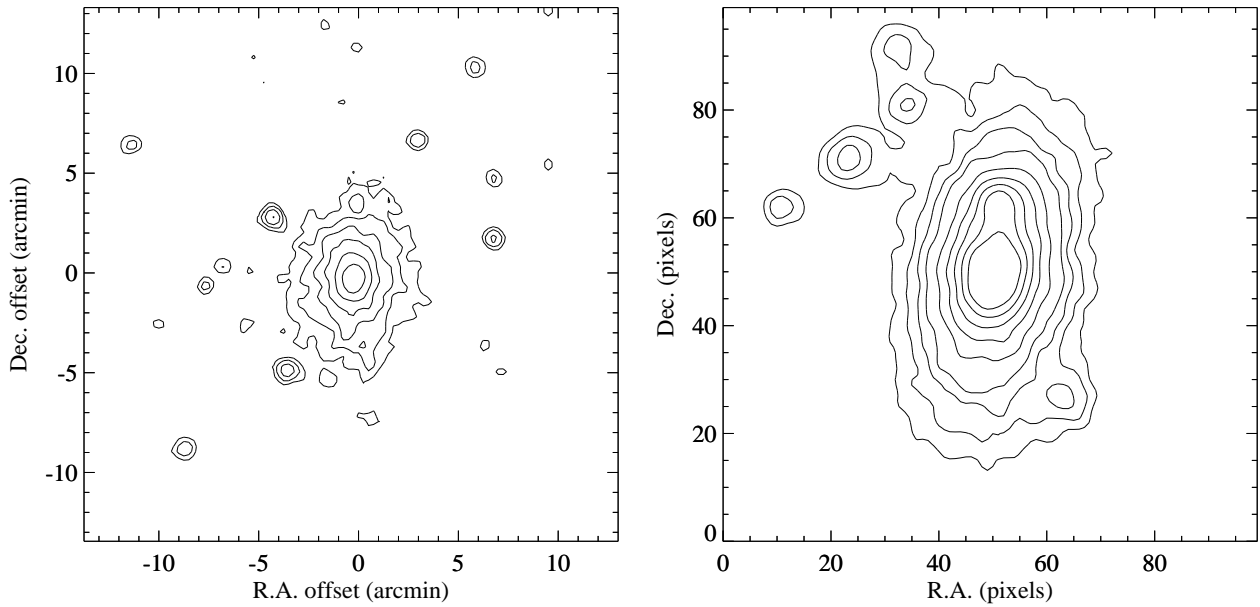

**Figure 3.** (a) Contours of constant X-ray surface brightness for Abell 1413. Details as in Fig. 1a. (b) Contours of constant optical surface brightness for the CCG of Abell 1413. Details as in Fig. 1b.

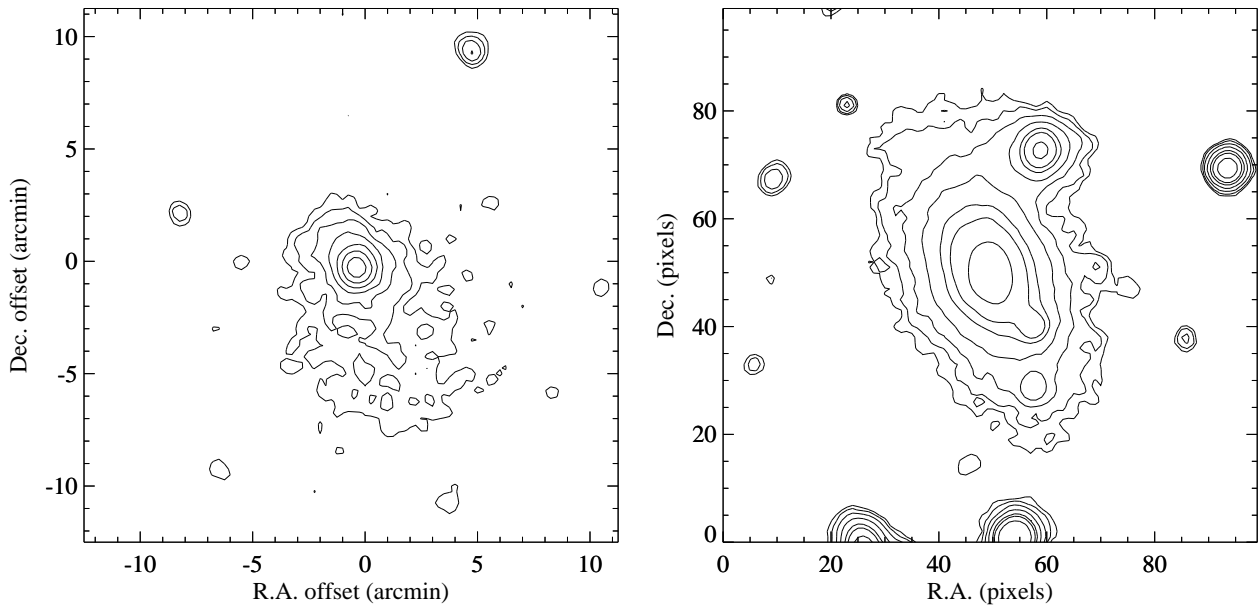

**Figure 4.** (a) Contours of constant X-ray surface brightness for Abell 1664. Details as in Fig. 1a. (b) Contours of constant optical surface brightness for the CCG of Abell 1664. The image was obtained with the 2.5 m Isaac Newton Telescope (JKT), La Palma in 1994 Jun and has been smoothed with a gaussian of FWHM 2 pixels. The spatial scale is 0.59 arcsec pixel$^{-1}$. For reference 10 pixels $\sim$ 6 arcsec, corresponding to $\sim$ 20 kpc at the redshift of Abell 1664.

**Table 1.** Summary of the X-ray observations

| Cluster | $z$ | Observation Date | Exp. | Exp. (corr.) |
|---|---|---|---|---|
| Abell 1068 | 0.1386 | 1992 Nov 30 | 10648 | 9616 |
| * Abell 1361 | 0.1167 | 1993 Jun 07 | 5675 | 3721 |
| # Abell 1413 | 0.1427 | 1991 Nov 27 | 11363 | — |
| Abell 1664 | 0.1276 | 1992 Jul 23 | 12766 | 12369 |
| Abell 2208 | 0.1329 | 1993 Mar 14 | 10663 | 7696 |

Notes: A summary of the X-ray observations. Column (1) lists the cluster name and (2) the redshift of the CCG from Allen *et al.* (1992). Column (3) gives the date of the observation. In Column (4) we list the raw exposure times in seconds and in (5) the exposures after corrections for satellite dead time and the removal of periods of high particle background and scattered solar X-ray contamination. * For Abell 1361 the corrected exposure has not been corrected for dead time (a ∼ 2 per cent effect). # The X-ray image of Abell 1413 was extracted from the ROSAT Data Archive (Böhringer P.I.). The redshift is from Struble & Rood (1987). No corrections for dead time, scattered solar X-rays or particles were made to the Abell 1413 data.

**Table 2.** The X-ray cluster morphology

| Name | X-ray Peak (2000.) | | Ellipticity | | PA | |
|---|---|---|---|---|---|---|
| | R.A. | Dec. | $0 - 0.5$ Mpc | $0.5 - 1$ Mpc | $0 - 0.5$ Mpc | $0.5 - 1$ Mpc |
| Abell 1068 | $10^h40^m45^s$ | $39°57'10''$ | $0.22 \pm 0.02$ | $0.25 \pm 0.04$ | $130 \pm 3$ | $148 \pm 5$ |
| Abell 1361 | $11^h43^m39^s$ | $46°21'10''$ | $0.22 \pm 0.03$ | — | $140 \pm 4$ | — |
| Abell 1413 | $11^h55^m19^s$ | $23°24'10''$ | $0.28 \pm 0.02$ | $0.27 \pm 0.02$ | $175 \pm 3$ | $179 \pm 2$ |
| Abell 1664 | $13^h03^m42^s$ | $-24°14'50''$ | $0.19 \pm 0.03$ | $0.30 \pm 0.05$ | $24 \pm 5$ | $30 \pm 7$ |
| Abell 2208 | $16^h29^m42^s$ | $58°31'30''$ | $0.34 \pm 0.04$ | $0.20 \pm 0.05$ | $172 \pm 4$ | $147 \pm 9$ |

Notes: A summary of the results from the isophote analysis of the PSPC cluster data. Columns (2) and (3) list the right ascension and declination of the peak of the X-ray emission. (The error in the aspect solution for a typical PSPC observation is ∼ 15 arcsec.) Columns (4) and (5) give the mean ellipticity $(1 - b/a)$ within 0.5 Mpc and between 0.5 and 1 Mpc respectively. Columns (6) and (7) give the mean position angle (PA) in the same intervals. For A1361 insufficient counts were obtained in the $0.5 - 1$ Mpc annulus to provide reliable ellipticity and PA estimates.

with the Galactic value determined from HI observations (Stark *et al.* 1992) was assumed to lie (at zero redshift) in the line of sight to the cluster. The detailed results from the deprojection analyses are presented in Figs. 6–10. In Table 4 we summarize the cooling flow results and list the cooling times in the central 30 arcsec (∼ 100 kpc) bins, the cooling radii (*i.e.* those radii within which the mean emission-weighted cooling time of the gas is less than the Hubble time; $H_0^{-1} = 2 \times 10^{10}$ yr) and the integrated mass deposition rates within the cooling radii.

The deprojection results show that all three of the optical line-emitting clusters (Abell 1068, Abell 1361 and Abell 1664) contain large cooling flows (of 400, 190 and 260 $M_\odot$ yr$^{-1}$ respectively). The central cooling times in these clusters are all very short, ∼ $2 \times 10^9$ yr, and in each case the mass deposition is concentrated within the inner 100 kpc. Abell 1413, which has no emission lines, also has a large cooling flow of 200 $M_\odot$ yr$^{-1}$, although the mass deposition is less centrally concentrated than in the three line-emitting clusters. The central cooling time is also significantly longer, ∼ $8 \times 10^9$ yr. The data for Abell 2208 (no emission lines) are consistent with no cooling flow and the central cooling time is $1.3^{+0.8}_{-0.4} \times 10^{10}$ yr. [Note, however, that when interpreting results on central cooling times it is important to remember that the values depend on the spatial resolution of the detector. Studies of nearby cooling flows indicate that the central cooling time reduces as the spatial resolution increases and smaller regions are resolved (*e.g.* see Edge, Stewart & Fabian 1992)].

The highly disturbed morphology in the outer regions of Abell 1664 does not severely affect the deprojection results. The mass deposition in this cluster is concentrated within the inner ∼ 200 kpc where the assumption of spherical symmetry is reasonable. Undoubtedly the major uncertainty in the deprojection analyses is in the assumed form for the gravitational potentials of the clusters. Simulations in which we have examined a wide range of plausible potentials suggest that the integrated mass deposition rates quoted in Table 4 should be regarded as uncertain by 30 − 50 per cent.

## 5 SPECTRAL ANALYSIS

A simple spectral analysis of the PSPC data has been carried out. For Abell 1361 and Abell 2208 only 1000 − 2000 counts were detected in the observations and, for these clusters, single integrated cluster spectra were extracted. For Abell 1068 and Abell 1664 sufficient counts were obtained to construct spectra in two concentric annuli of radii $0 - 2$ arcmin and $2 - 6$ arcmin, and to thereby investigate (to first order) any strong radial variation in the spectral properties of the

**Table 3.** The optical CCG morphology

| Name | Abell Position (2000.) | | CCG Position (2000.) | | Ellipticity | PA |
|---|---|---|---|---|---|---|
| | R.A. | Dec. | R.A. | Dec. | | |
| Abell 1068 | $10^h40^m48^s$ | $39°57'$ | $10^h40^m44^s$ | $39°57'10''$ | $0.32 \pm 0.03$ | $129 \pm 3$ |
| Abell 1361 | $11^h43^m48^s$ | $46°21'$ | $11^h43^m39^s$ | $46°21'10''$ | $0.19 \pm 0.04$ | $136 \pm 6$ |
| #Abell 1413 | $11^h55^m24^s$ | $23°22'$ | $11^h55^m20^s$ | $23°24'7''$ | $0.38 \pm 0.01$ | $172 \pm 1$ |
| Abell 1664 | $13^h03^m42^s$ | $-24°13'$ | $13^h03^m42^s$ | $-24°14'43''$ | $0.40 \pm 0.03$ | $26 \pm 3$ |
| Abell 2208 | $16^h29^m36^s$ | $58°30'$ | $16^h29^m39^s$ | $58°31'51''$ | $0.37 \pm 0.02$ | $154 \pm 2$ |

Notes: A summary of the results from the isophote analysis of the optical CCG data. Columns (2) and (3) list the Abell, Corwin & Olowin (1989) cluster coordinates. Columns (4) and (5) list the CCG coordinates from Allen *et al.* 1992. Columns (6) and (7) give the mean ellipticity $(1 - b/a)$ and position angle (PA) between radii of 2 and 10 arcsec ($\sim 6 - 30$ kpc). # For Abell 1413 the CCG coordinates are from the Huchra redshift catalogue.

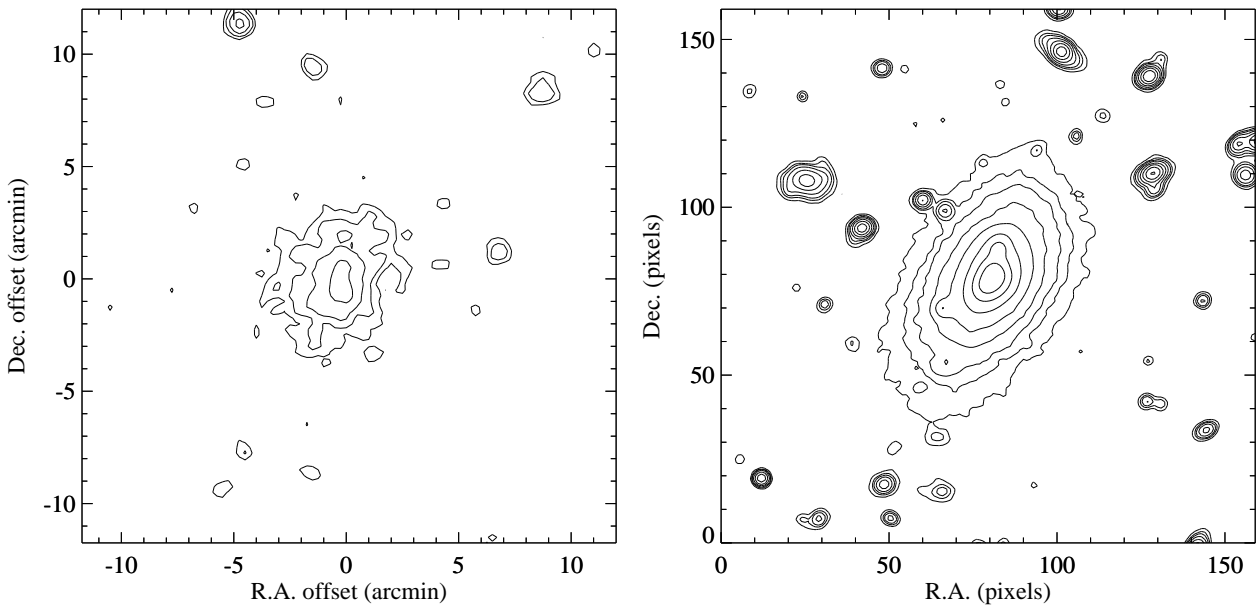

**Figure 5.** a) Contours of constant X-ray surface brightness for Abell 2208. Details as in Fig. 1a. (b) Contours of constant optical surface brightness for the CCG of Abell 2208. Details as in Fig. 4b.

clusters. Background spectra were extracted from circular regions of radius 0.1 deg, offset by $\sim 40$ arcmin from the field centres, which were free of contamination by strong foreground sources. The background spectra were scaled to, and subtracted from, the cluster spectra. The small unvignetted particle background was accounted for using the parametrization of Snowden *et al.* (1992).

The spectral analysis has been carried out using the XSPEC spectral fitting package (Shafer *et al.* 1991). The PSPC spectra were binned-up by a factor 8 in PHA space, and then further binned to ensure a minimum of 20 counts per PHA channel, enabling a reliable $\chi^2$ analysis to be carried out. PHA channels corresponding to energies below 0.1 keV (which are not correctly calibrated by the PSPC response matrix) and above 2.4 keV (where the particle background dominates) were ignored. The most recent version of the ROSAT PSPC response matrix (Version 36) has been used in the analysis. A systematic error of 1 per cent of the measured count rate in each PHA channel has been quadratically summed with the photon noise in those channels to allow for uncertainties in the calibration of the PSPC response matrix.

The spectra were first fitted with simple, single temperature models. An updated version of the code of Raymond & Smith (1977) for X-ray emission from hot, diffuse, thermal plasmas was used, together with the photoelectric absorption models of Morrison & McCammon (1983). Redshifts were fixed at the values determined optically for the CCGs (Allen *et al.* 1992). The free parameters in the fits were the temperature, metallicity, and emission measure of the X-ray gas and the absorbing column density. The best-fit $\chi^2$

**Table 4.** Cooling flow results

| Name | $\sigma_r$ | $N_H$ | $r_{cool}$ | $t_{cool}$ | $\dot{M}$ |
|------|-----------|-------|-----------|-----------|-----------|
| Abell 1068 | 900 | 0.16 | $220^{+30}_{-70}$ | $2.0^{+0.1}_{-0.1}$ | $400^{+20}_{-10}$ |
| Abell 1361 | 800 | 0.22 | $200^{+100}_{-70}$ | $2.3^{+0.3}_{-0.3}$ | $190^{+70}_{-40}$ |
| Abell 1413 | 1000 | 0.20 | $180^{+80}_{-30}$ | $8.3^{+0.5}_{-0.7}$ | $200^{+80}_{-30}$ |
| Abell 1664 | 900 | 0.86 | $190^{+40}_{-50}$ | $2.7^{+0.2}_{-0.2}$ | $260^{+20}_{-20}$ |
| Abell 2208 | 670 | 0.18 | $150^{+90}_{-150}$ | $13^{+8}_{-4}$ | $60^{+40}_{-60}$ |

Notes: A summary of the results on the cooling flows from the deprojection analyses of the X-ray images. Column (2) lists the estimated cluster velocity dispersions (in km s$^{-1}$). Column (3) lists the Galactic column densities in units of $10^{21}$ atom cm$^{-2}$ (Stark *et al.* 1992). Cooling radii ($r_{cool}$) are quoted in kpc, cooling times for the central 30 arcsec bin ($t_{cool}$) in $10^9$ yr, and mass deposition rates ($\dot{M}$) in M$_\odot$ yr$^{-1}$. Errors on the cooling times describe the 10 and 90 percentile values from 100 Monte Carlo simulations. The upper and lower confidence limits on the cooling radii mark the radii where the 10 and 90 percentiles exceed, and become less than, the Hubble time respectively. Errors on the mass deposition rates are the 90 and 10 percentile values at the upper and lower limits for the cooling radius.

**Table 5.** Results from the X-ray spectra

| | | 0 − 2 arcmin | | | | 2 − 6 arcmin | | |
|------|------|------|------|---------|------|------|------|---------|
| Name | $kT$ | $N_H$ | $Z$ | $\chi^2$/DOF | $kT$ | $N_H$ | $Z$ | $\chi^2$/DOF |
| Abell 1068 | $2.3^{+0.6}_{-0.4}$ | $0.16^{+0.03}_{-0.02}$ | $0.38^{+0.33}_{-0.20}$ | 21.9/22 | $5.6^{+\infty}_{-2.3}$ | $0.18^{+0.15}_{-0.10}$ | UC | 12.7/18 |
| Abell 1361* | — | — | — | — | $1.8^{+1.2}_{-0.4}$ | $0.31^{+0.10}_{-0.10}$ | $0.2^{+0.6}_{-0.2}$ | 27.4/22 |
| Abell 1664 | $2.2^{+1.3}_{-0.8}$ | $1.22^{+0.55}_{-0.28}$ | $0.35^{+1.06}_{-0.30}$ | 15.8/21 | $6.5^{+\infty}_{-4.9}$ | $1.39^{+1.28}_{-0.52}$ | UC | 14.3/17 |
| Abell 2208* | — | — | — | — | $3.2^{+2.7}_{-1.1}$ | $0.12^{+0.09}_{-0.08}$ | UC | 11.2/23 |

| | | Cooling flow model | | | |
|------|------|--------|------|------|---------|
| Name | $\dot{M}$ | $\Delta N_H$ | $F_X$ | $L_X$ | $\chi^2$/DOF |
| Abell 1068 | 400 | $> 0.9$ | $7.6 \pm 0.2$ | $11.0 \pm 0.3$ | 30.6/22 |
| Abell 1361 | 190 | UC | $4.4 \pm 0.1$ | $3.7 \pm 0.1$ | 28.4/22 |
| Abell 1664 | 260 | $1.2^{+2.1}_{-0.7}$ | $5.7 \pm 0.2$ | $7.4 \pm 0.2$ | 19.6/22 |
| Abell 2208 | — | — | $2.3 \pm 0.1$ | $2.6 \pm 0.1$ | — |

Notes: A summary of best-fit parameters and confidence limits from the X-ray spectral analysis. Columns $1 - 9$ describe the results from the fits with the simple, single-temperature models. Temperatures ($kT$) are expressed in keV, total (Galactic plus intrinsic) column densities ($N_H$) in $10^{21}$ atom cm$^{-2}$ and metallicities ($Z$) as a fraction of the solar value. Columns $10 - 14$ describe the results for the cooling-flow models fitted to integrated cluster spectra. The $\Delta N_H$ values in column 11 are the measured excess absorptions over the Galactic values. X-ray fluxes ($F_X$) are in units of $10^{-12}$ erg cm$^{-2}$ s$^{-1}$ and absorption-corrected X-ray luminosities ($L_X$) are in $10^{44}$ erg s$^{-1}$. Both ($F_X$) and ($L_X$) are quoted in the observed $0.1 - 2.4$ keV ROSAT band. All confidence limits are quoted at the 90 per cent level ($\Delta\chi^2 = 2.71$) for a single interesting parameter. UC indicates that a parameter is unconstrained at the 90 per cent confidence level. *For Abell 1361 and Abell 2208 the results listed in the $2 - 6$ arcmin annulus column are the results for single temperature fits to the integrated spectra from $0 - 6$ arcmin.

and parameter values and the 90 per cent confidence limits on the parameters are summarised in Table 5.

The results of the deprojection analysis described in Section 4 show that Abell 1068, Abell 1361 and Abell 1664 contain large cooling flows. If the emitting gas is substantially multiphase (as it will be in the case of the projected spectrum of a large cooling flow) the variation of emissivity with temperature and the limited $0.1 - 2.4$ keV bandpass of the PSPC will result in measured temperatures being strongly weighted by the lower temperature components. It is important, therefore, to remember that the best-fit temperatures listed in Table 5 are emission-weighted values (particularly in the case of Abell 1361, for which count statistics were only sufficient for a single, integrated spectrum of the cluster to be formed).

We next examined integrated spectra for the cooling flow clusters using a more physical model in which a component was included to account for the emission from gas cooling to zero temperature at constant pressure (*i.e.* a constant pressure cooling flow) from an assumed ambient cluster temperature of 6 keV. The nor-

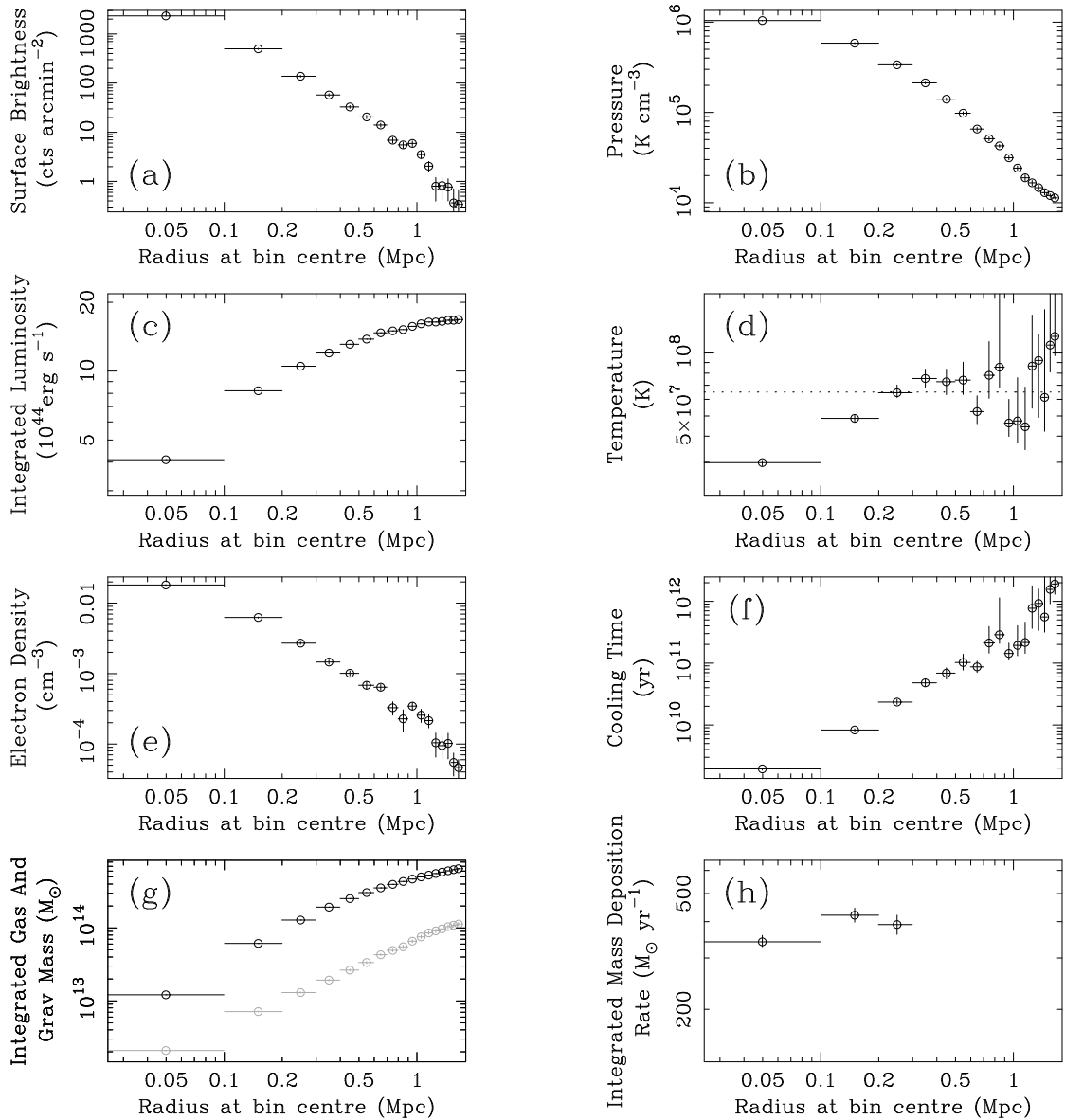

**Figure 6.** A summary of the results from the deprojection analysis of Abell 1068. From left to right, top to bottom, we plot; (a) surface brightness, (b) pressure, (c) integrated luminosity, (d) temperature, (e) electron density, (f) cooling time, (g) integrated gas and gravitational mass and (h) integrated mass deposition rate. In Figs. (a), (b), (c), (e) and (g) the plotted points show the mean values and $1\sigma$ errors (in each radial bin) from 100 Monte Carlo simulations. In Figs. (d), (f) and (h) the points and errors mark the median and 10 and 90 percentile values from 100 Monte Carlo simulations. The dashed line in (c) marks the best-fit temperature using the spectral data for the $2-6$ arcmin annulus (where the emission-weighting effects of the cooling flow may be largely ignored).

malization of this component was fixed at the mass deposition rate determined from the deprojection analysis. Studies of cooling flows in nearby clusters observed with the Solid State Spectrometer (White *et al.* 1991) and ROSAT PSPC (Allen *et al.* 1993) have demonstrated the presence of large masses of intrinsic, centrally concentrated X-ray absorbing material. To test for the presence of intrinsic absorption in the BCS clusters, we allowed the column density acting on the cooling flows to be a free parameter in the fits. [The column acting on the ambient cluster gas was maintained at

the Galactic value determined from HI measurements (Stark *et al.* 1992; see Table 4).]

The results of the fits with the cooling-flow models are summarized in Table 5. The spectral data for all three cooling-flow clusters are well-described by these models (although the inclusion of the cooling-flow components does not significantly improve the $\chi^2$ values over the single-temperature fits). Using the cooling-flow models we find evidence for excess absorption in two of the cooling-flow clusters. The column density acting on the cooling flow in Abell 1664 is measured to be $1.2^{+2.1}_{-0.7} \times 10^{21}$ atom $cm^{-2}$ above the Galactic value. For Abell 1068 the excess absorption is determined to be $> 0.9 \times 10^{21}$ atom $cm^{-2}$ (all limits quoted at 90 per cent confidence). No useful constraint on intrinsic absorption in the cooling flow of Abell 1361 was obtained. We note that no correction for the redshift dependence of column density [approximately $N_H(z) \propto (1 + z)^3$] has been applied to the above values. The intrinsic absorbing material has been modelled as a uniform screen, local to the observer, and the true column density of the absorbing material in the clusters should therefore be $\sim 40$ per cent greater than the measured values. (The redshift-corrected intrinsic column densities are listed in Table 6.) It is also important to note that the projected spectrum of a real, intrinsically absorbed cooling flow will undoubtedly be more complicated than the simple models used here.

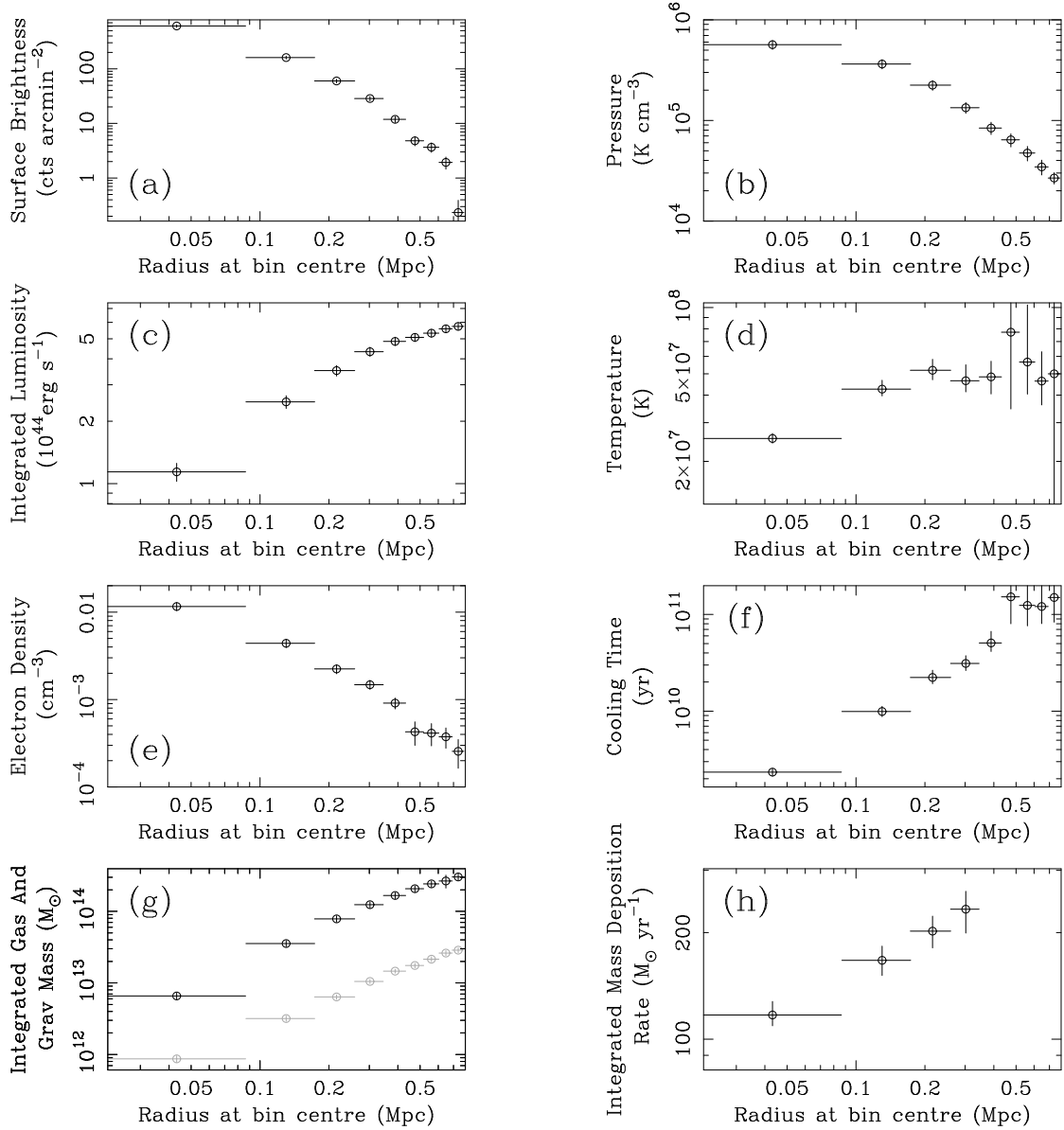

**Figure 7.** Deprojection results for Abell 1361. Details as in Fig. 6. No useful constraint on the ambient cluster gas temperature was obtained from the spectral data.

## 6  DISCUSSION

### 6.1  Cooling flows and optical emission lines

We have shown that three of the most optically line-luminous clusters (*i.e.* those clusters with the most optically line-luminous CCGs), selected from the $0.1 < z < 0.15$ interval of the ROSAT BCS, all contain large cooling flows and exhibit short central cooling times. In contrast, both Abell 1413 and Abell 2208, which do not exhibit optical emission lines, have longer central cooling times (although the integrated mass deposition rate for Abell 1413 is similar to those of the line-emitting clusters). Although no simple correlation between $\dot{M}$, $t_{cool}$ and optical line luminosity exists (other well-studied clusters such as Abell 2029 and Abell 1689 also have substantial cooling flows but no detectable optical line emission) there is a clear tendency for the most optically-line-luminous central galaxies to be in clusters with large cooling flows and short central cooling times. Note also that although the line-emitting CCGs in our sample exhibit a range of optical emission-line ratios (suggesting different ionization states in the line emitting gas) we find no obvious differences in the X-ray spectral properties of their host clusters.

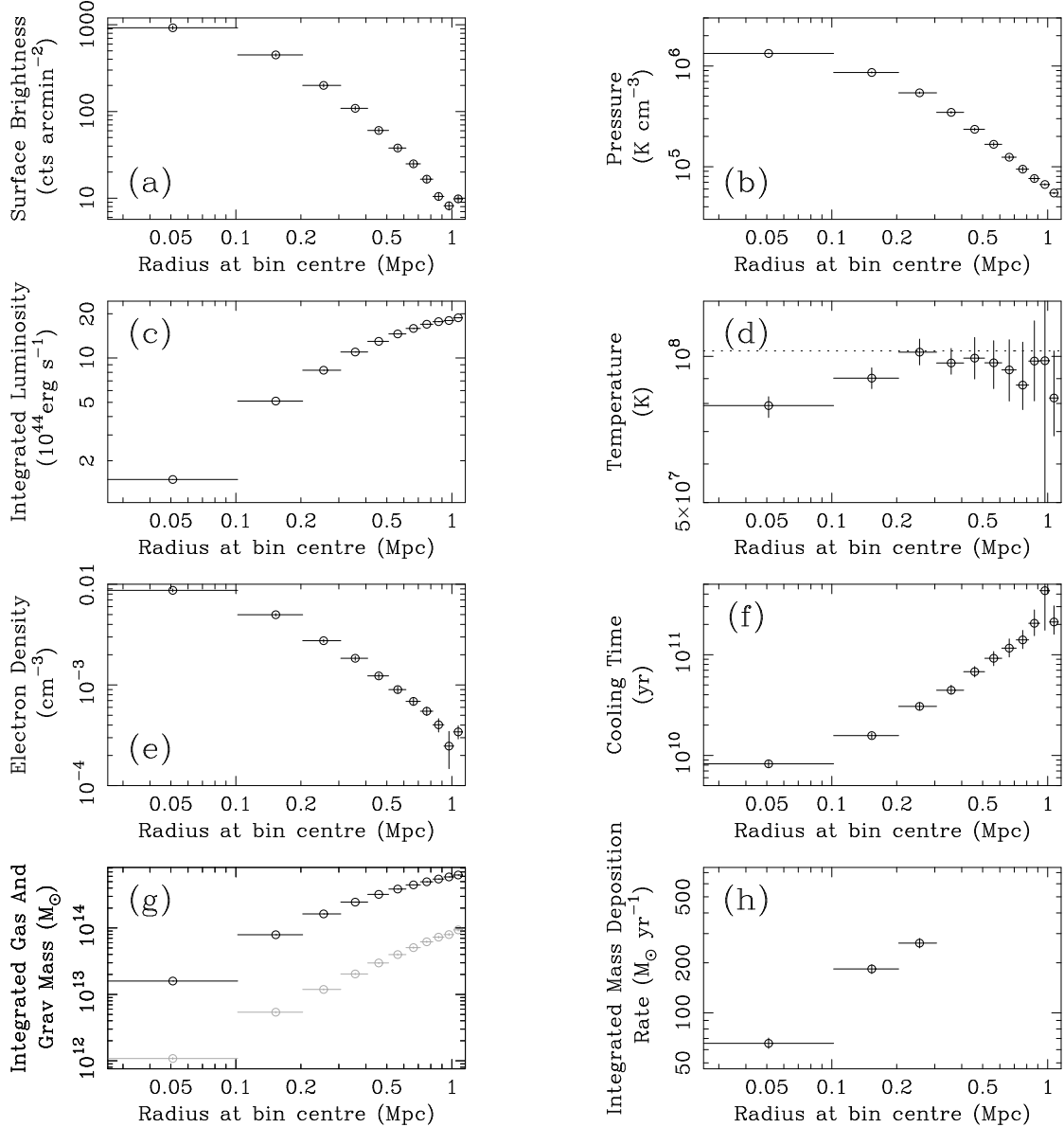

**Figure 8.** Deprojection results for Abell 1413. Details as in Fig. 6. The spectral temperature constraint is from David *et al.* 1993.

### 6.2 X-ray absorption and intrinsic reddening

The combination of X-ray image deprojection and spectral results indicate that the cooling flows in Abell 1068 and Abell 1664 are intrinsically absorbed. The level of intrinsic absorption observed ($\gtrsim 10^{21}$ atom cm$^{-2}$) is similar to that observed in clusters at lower redshift (White *et al.* 1991; Allen *et al.* 1993) although the current data alone cannot place firm constraints on the spatial distribution and mass of the absorbing material. For illustration though, a uni-form sheet of absorbing material with a column density of $2 \times 10^{21}$ atom cm$^{-2}$ (and cosmic abundances) acting within a cooling radius of $\sim 200$ kpc would imply a mass of absorbing material of $\sim 2 \times 10^{12}$ M$_\odot$. Such a mass could have been accumulated by a cooling flow of 200 M$_\odot$ yr$^{-1}$ in $\sim 10^{10}$ yr.

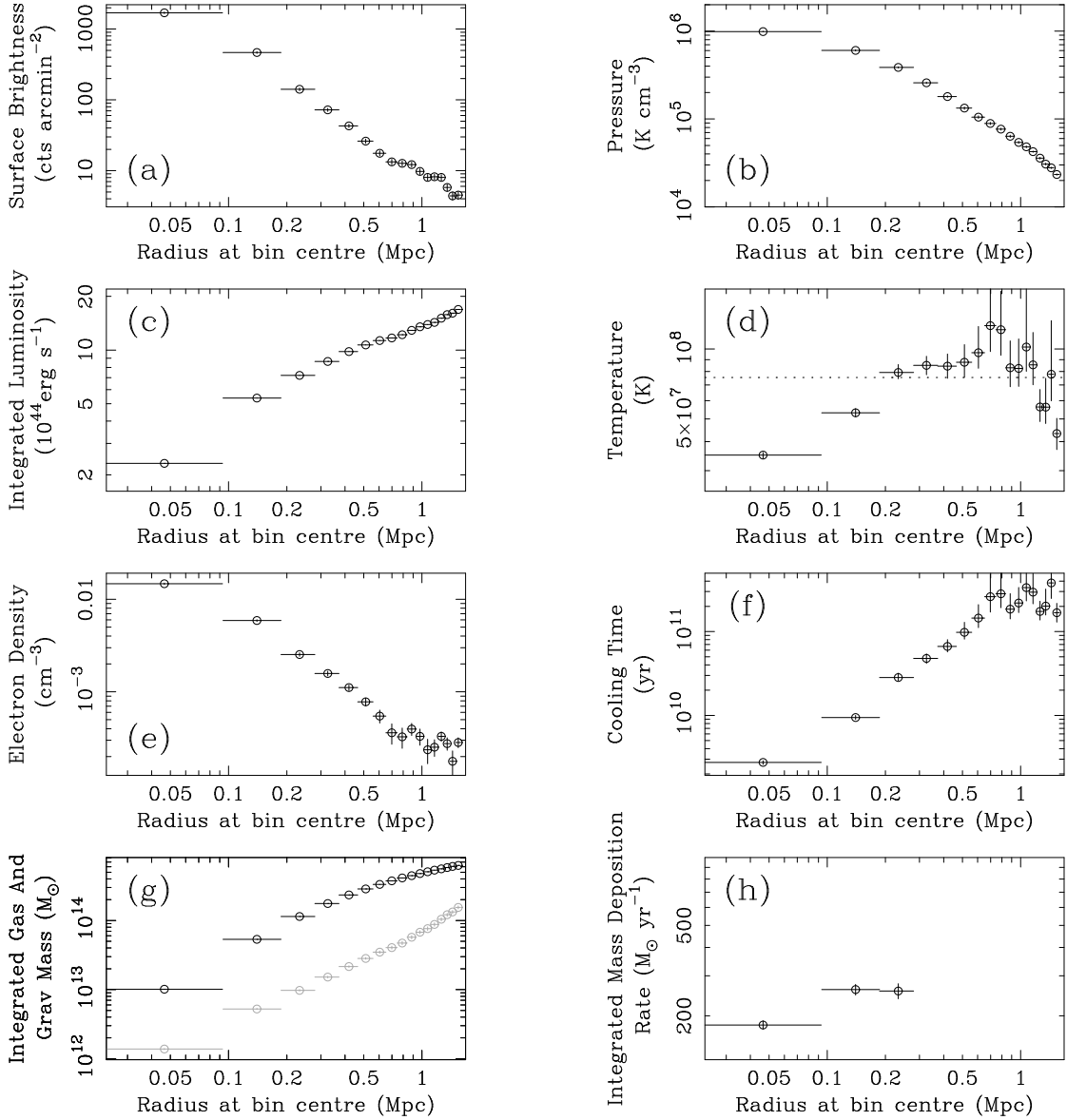

**Figure 9.** Deprojection results for Abell 1664. Details as in Fig. 6.

An interesting comparison can be made between the column densities inferred from the X-ray data and the reddening implied by the optical H$\alpha$/H$\beta$ emission line ratios. Using the line fluxes given in Allen *et al.* (1992) we find that that for Abell 1068 H$\alpha$/H$\beta = 5.6^{+1.8}_{-1.3}$ and for Abell 1664 H$\alpha$/H$\beta = 6.2^{+1.7}_{-1.2}$ [a preliminary correction for Galactic reddening has been made in each case; errors on the H$\alpha$/H$\beta$ ratios are the maximum and minimum values within the upper and lower 90 per cent confidence flux limits on H$\alpha$ and H$\beta$ from Allen *et al.* (1992)]. If we assume that the differences between the observed H$\alpha$/H$\beta$ values and simple Case B recombination (H$\alpha$/H$\beta \sim 2.86$ for gas at $10^4$ K; Ferland & Osterbrock 1985) are due to reddening by material intrinsic to the clusters, we derive values for this intrinsic reddening of $E(B-V) \sim 0.54^{+0.23}_{-0.21}$ for Abell 1068 and $E(B-V) \sim 0.63^{+0.19}_{-0.17}$ for Abell 1664. [The reddening laws of Seaton (1979) and Howarth (1983) have been used]. Applying the standard (Galactic) transformation between colour excess and column density in neutral hydrogen, $N_H/E(B-V) = 5.8 \times 10^{21}$ atom cm$^{-2}$ mag$^{-1}$ (Bohlin, Savage & Drake 1978), we then infer intrinsic neutral hydrogen column densities in the line of sight to the CCGs of Abell 1068 and Abell 1664 of $3.1^{+1.3}_{-1.2}$ and $3.7^{+1.1}_{-1.0} \times 10^{21}$ atom cm$^{-2}$ respectively.

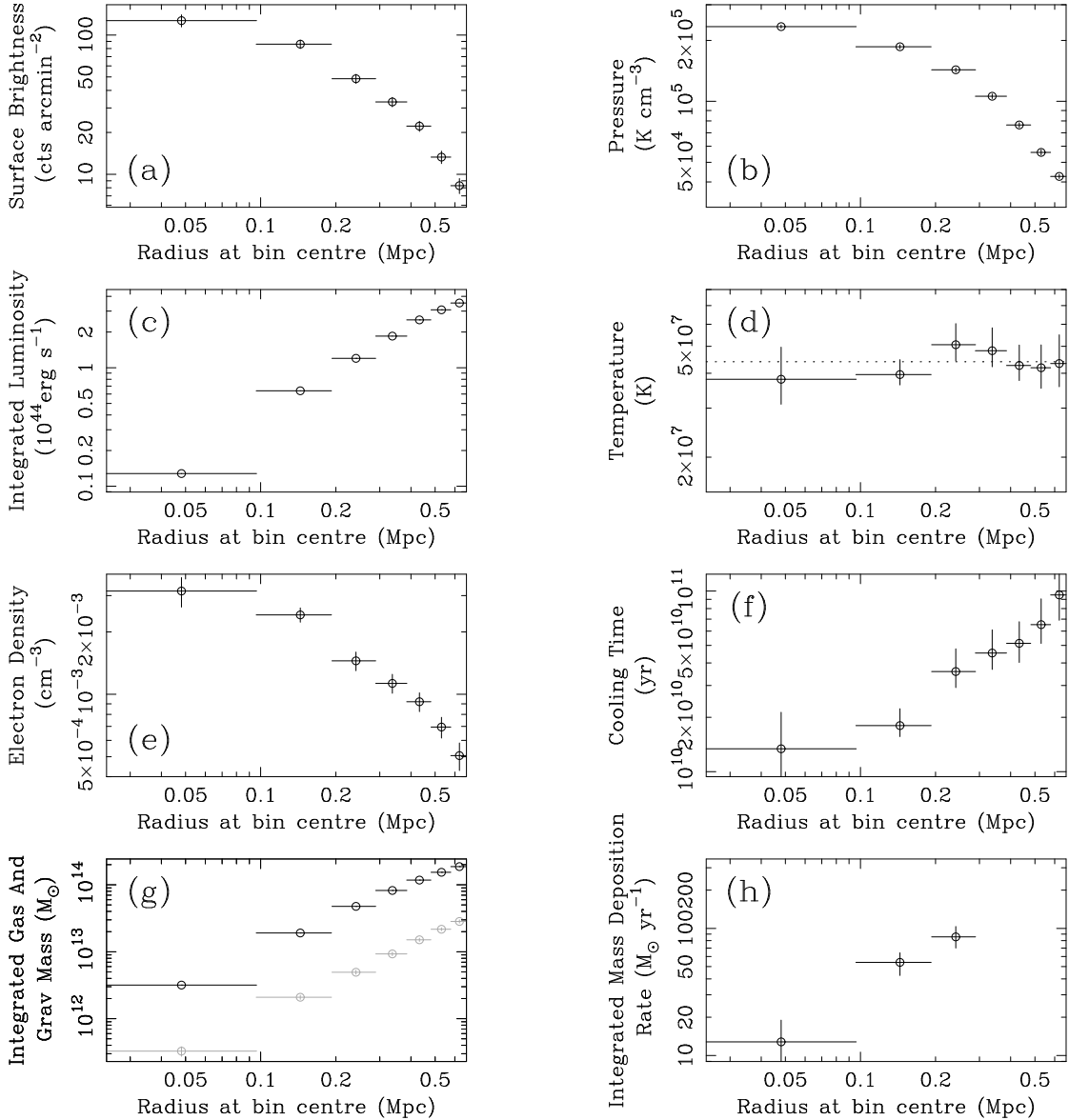

**Figure 10.** Deprojection results for Abell 2208. Details as in Fig. 6.

We should note that several uncertainties are present in deriving such absorption estimates. Firstly there are no *a priori* reasons why the standard Galactic gas/dust ratio should apply to the conditions at the centre of a large cooling flow. Secondly, Case B recombination may not strictly apply, although for the photoionization models for cooling-flow nebulae discussed by Crawford & Fabian (1992) and the stellar photoionization models discussed by Allen (1995), the Case B approximation is a fair one. [Note that the earlier models of Donahue & Voit (1991) predict un-reddened Hα/Hβ ratios ∼ 4.] Evidence supporting the Case B approximation is also found in the agreement of the reddening estimates inferred from the Hγ/Hβ and Hα/Hβ ratios for Abell 1068. The observed Hγ/Hβ ratio for this cluster of 0.37 implies the same intrinsic reddening, $E(B-V) = 0.54$, as the Hα/Hβ ratio. (The signal-to-noise in the other CCG spectra are insufficient to allow reliable Hγ flux measurements to be made.)

In Table 6 we compare the intrinsic column densities inferred from the optical data to those observed in the PSPC X-ray spectra. Also tabulated are the $E(B-V)$ and implied $\Delta N_H$ values, and the X-ray column densities for the five clusters common to the UV/optical study of Hu (1992) and the X-ray study of White *et al.* (1991). One should note the aperture sizes used in the various

**Table 6.** A comparison of intrinsic absorption estimates from the optical, UV and X-ray data

| Name | $z$ | OPTICAL | | | UV | | | X-RAY |
|---|---|---|---|---|---|---|---|---|
| | | $H\alpha/H\beta$ | $E(B-V)$ | $\Delta N_H$ | $Ly\alpha/H\alpha$ | $E(B-V)$ | $\Delta N_H$ | $\Delta N_H$ |
| A1068 | 0.1386 | $5.6^{+1.8}_{-1.3}$ | $0.54^{+0.23}_{-0.21}$ | $3.1^{+1.3}_{-1.2}$ | NA | UC | UC | $>1.3$ |
| A1664 | 0.1276 | $6.2^{+1.7}_{-1.2}$ | $0.63^{+0.19}_{-0.17}$ | $3.7^{+1.1}_{-1.0}$ | NA | UC | UC | $1.7^{+3.0}_{-1.0}$ |
| A1361 | 0.1167 | $>2.4$ | $>-0.3$ | $>-1.7$ | NA | UC | UC | UC |
| A2208 | 0.1329 | — | UC | UC | NA | UC | UC | — |
| A1795 | 0.0633 | $>3.7$ | $>0.22$ | $>1.3$ | 6.1 | $0.14^{+0.08}_{-0.12}$ | $0.8^{+0.5}_{-0.7}$ | $0.8^{+0.3}_{-0.3}$ |
| A85 | 0.0518 | NA | NA | NA | $<10.4$ | $0.04^{+0.07}_{-0.04}$ | $0.2^{+0.4}_{>0.2}$ | $1.0^{+0.5}_{-0.4}$ |
| A426 | 0.0183 | NA | NA | NA | 1.5 | $0.24^{+0.08}_{-0.10}$ | $1.4^{+0.5}_{-0.6}$ | $1.3^{+0.2}_{-0.2}$ |
| A496 | 0.0320 | NA | NA | NA | 2.8 | $0.23^{+0.07}_{-0.09}$ | $1.3^{+0.4}_{-0.5}$ | $2.0^{+1.1}_{-0.2}$ |
| A2199 | 0.0309 | NA | NA | NA | 2.8 | $0.25^{+0.07}_{-0.09}$ | $1.5^{+0.4}_{-0.6}$ | $1.4^{+0.2}_{-0.2}$ |

Notes: A comparison of $E(B-V)$ estimates derived from the relative strengths of the H$\alpha$, H$\beta$ and Ly$\alpha$ lines with the intrinsic absorption ($\Delta N_H$) inferred from the X-ray data. All $\Delta N_H$ values are quoted in units of $10^{21}$ atom cm$^{-2}$ and have been corrected for redshift effects.. The conversion between $E(B-V)$ and $\Delta N_H$ has been carried out using the relation of Bohlin, Savage & Drake (1978). The optical data are described by Allen *et al.* (1992). The UV observations and Ly$\alpha$/H$\alpha$ are from Hu (1992). (Note that the limits on the Hu values are not error bars but reflect systematic uncertainties in the intrinsic Ly$\alpha$/H$\alpha$ values. Hu estimates errors in the measurement of the Ly$\alpha$/H$\alpha$ ratio to be $\sim$ 30 per cent.) Intrinsic X-ray absorption estimates for Abell 1795, Abell 85, Abell 426, Abell 496 and Abell 2199 are from White *et al.* (1991).

observations. The optical CCG studies detailed in Allen *et al.* (1992) were carried out using a slit of $1.6 \times 6$ arcsec$^2$ [corresponding to a spatial scale of $5 \times 18$ kpc$^2$ at a redshift $z = 0.125$. Note that no aperture correction is involved in determining the H$\alpha$/H$\beta$ ratios from the optical spectra and that these observations were all made at the parallactic angle, thereby minimizing differential atmospheric dispersion effects]. In the spectral fits to the PSPC data (Section 5) the intrinsic absorption was assumed to act only on the cooling flow. The deprojection analysis of Section 4 shows that in both Abell 1068 and Abell 1664 the cooling material is deposited within the central 200 kpc of the clusters (see Figs. 6,9). The ultraviolet observations of Hu (1992) were made using the large $(8.9 \times 21.6$ arcsec$^2$; corresponding to a spatial scale of $\sim 10 \times 24$ kpc$^2$ at the mean redshift of these observations of $z \sim 0.04$) aperture on the International Ultraviolet Explorer. Narrow band H$\alpha$ images of the same spatial regions were used to obtain the quoted Ly$\alpha$/H$\alpha$ ratios. The White *et al.* (1991) X-ray observations were made with the Solid State Spectrometer (SSS) on the Einstein Observatory which had a 3 arcmin radius ($\sim 200$ kpc at $z = 0.04$) circular field of view.

The X-ray, optical and UV data all indicate the presence of large amounts of intrinsic absorbing material in the central regions of the clusters. The agreement between the column densities measured across the wavebands and over a range of apertures suggests that the bulk of the material responsible for the intrinsic reddening may be associated with the X-ray absorbing gas. In addition, the results imply that the properties of dust, and the dust-to-gas ratios, in the X-ray absorbing gas are similar to those of material in our own Galaxy. [Note that a similar correlation between $E(B-V)$ and $\Delta N_H$ is also observed in the Centaurus cluster (Allen & Fabian 1994).] A theoretical discussion of dust formation in cooling flows is given by Fabian, Johnstone & Daines (1994). The implications of intrinsic reddening for our understanding of the UV/blue continuum in line-luminous CCGs is discussed by Allen (1995).

## 6.3   CCG–cluster alignments

One of the most immediately apparent results from the present study is the close correspondence between the optical CCG and cluster X-ray morphologies. As noted in Section 1, the tendency for CCGs to align with the axes of the optical galaxy distribution in their host clusters is well documented. The major advance of X-ray studies over the earlier optically-based work arises from the very short collision timescale of the intracluster gas in comparison to its dynamical and cooling timescales. (In contrast, the collision timescale for galaxies, other than the most massive galaxies in the cluster centre, generally exceeds the Hubble time.) Whereas the galaxy distribution can be complicated by the effects of subcluster mergers occurring after the initial cluster collapse, the X-ray ICM provides a tracer of the cluster potential as it evolves. [For further discussion of the distributions of the intracluster gas, galaxies, and dark matter in a sample of Abell clusters see Buote & Canizares (1992).]

The position angles of the $(0 - 0.5$ Mpc) cluster X-ray and optical CCG emission for Abell 1068, Abell 1361, Abell 1413 and Abell 1664 all agree to within 5 degree. The ellipticities of the CCG isophotes (averaged over radii of $2 - 10$ arcsec; corresponding to a physical scale of $6 - 32$ kpc for a redshift of 0.125) are typically $20 - 40$ per cent larger than those of the clusters (measured from the X-ray data). This is in agreement with the results of Porter *et al.* (1991) who demonstrated that the ellipticities of brightest cluster members in Abell clusters generally exceeds those of their host clusters at large radii. The CCGs of all 5 clusters lie at (or close to) the peak of the X-ray emission (the centroids of which are not observed to vary significantly with radius) implying that the CCGs rest at the base of the gravitational potentials of the clusters.

Clearly some mechanism is responsible for communicating the form of the cluster potential, defined by the X-ray emission on a scale of Mpc, directly to the CCG in the central few kpc. Models of hierarchical (cold dark matter) cluster collapse suggest that galaxies form well before the clusters undergo their initial collapse. The unique relationship between CCGs and their hosts, however, may suggest that CCGs form during or after the collapse of the cluster, or at least have been substantially modified since the cluster was formed.

A popular mechanism for CCG formation, discussed widely in the literature, is the 'galactic cannibalism' model (Ostriker & Tremaine 1975; Gunn & Tinsley 1976; White 1976; Ostriker & Hausman 1977). Dynamical friction forces are thought to cause the orbits of the most massive galaxies in a cluster to decay and merge, leading to the formation of a single supergiant galaxy which then subsumes smaller galaxies that pass through the cluster centre. The cannibalism model has several attractive features and can, for example, explain the large luminosities of CCGs in relation to their surrounding galaxies. Limitations to the model do, however, exist. In particular, deep imaging studies suggest that although mergers play some role in CCG evolution, the present growth rate of CCGs through mergers is low (typically $\sim 1 L^*$ per Hubble time. Merritt 1985; Lauer 1988; Merrifield & Kent 1991). As discussed by Merritt (1985), the growth rate through mergers should be a sensitive function of the velocity dispersion of the host cluster (with the rate $\propto \sigma^{-7} r_c^2$; where $\sigma$ is the velocity dispersion and $r_c$ is the core radius of the cluster). That fact that CCGs are not confined to low $\sigma$ clusters is interpreted as evidence against the continuous cannibalism model. If mergers are the dominant CCG formation mechanism, it seems that they are required to be far more prevalent at early epochs.

A second mechanism, that may play an important role in the formation of CCGs in rich clusters, is the merging of subclusters. When a subcluster containing its own CCG merges with a cluster, the CCGs of the two systems also merge on a timescale of a few $10^9$ years (Tremaine 1990). The merging system is thought to pass through a dumbbell stage before forming a single, large CCG. It is interesting to note that unless the impact parameter, and therefore angular momentum, for merging CCGs is small, the core of the post-merger cluster is likely to be oblate (and to be viewed partly edge on). If such a mechanism is dominant in CCG formation, however, it is not clear that the close correspondence between the CCG and cluster morphologies observed in the present sample of clusters (on scales of kpc to Mpc) would be maintained.

Numerical studies of the dissipationless collapse of cold dark matter clusters (Rhee & Roos 1990) have shown that the collapse of prolate, initially homogeneous ellipsoids, can lead to the transfer of information concerning the orientation of the collapsing system to the centre of the particle distribution. The orientation of the cluster is then transferred to the centre of the potential during the process of violent relaxation. Such results are seen to be robust against the effects of substructure in the collapsing clusters. Furthermore, such simulations also reproduce the observed tendency for the ellipticity of CCGs to increase with radius (Porter *et al.* 1991). However, whether cluster cores are predominantly oblate or prolate is uncertain at present.

The small sample of clusters presented here allow us to speculate on the relation between CCG/X-ray alignments, cooling flows and the growth of clusters. The present sample of clusters includes some of the largest known cooling flows and the alignment between the CCGs and X-ray gas in these systems is particularly good. The alignment in Abell 2208, the one cluster for which the X-ray data are consistent with no cooling flow, is significantly poorer than for the four large cooling flow clusters. The presence of a cooling flow seems to be the natural state of rich, relaxed clusters of galaxies. Edge, Stewart & Fabian (1992) demonstrate that between 70 and

90 per cent of the 50 X-ray brightest clusters contain cooling flows. The clusters with the largest flows tend to be the largest, most relaxed systems and the absence of a cooling flow appears to indicate that the cluster has either undergone a recent massive merger event [simulations by McGlynn & Fabian (1984) demonstrate that a cooling flow can be severely disrupted by the merger of two similarly sized subclusters] or that the potential of the cluster is still rapidly evolving (shown by the relative paucity of cooling flows in irregular, low X-ray luminosity clusters). For rich clusters then, poor CCG–cluster alignment and the absence of a cooling flow may be linked and together be indicators of merger activity.

The high quality X-ray images available with ROSAT allow the relationships between CCGs and their host clusters to be examined in more detail than ever before. Detailed comparison of X-ray observations for a large number of clusters with optical and UV observations of their CCGs will provide important clues to the dominant processes involved in the formation and evolution of central cluster galaxies. Several outstanding issues may be best-addressed by numerical simulation such as the effects of subcluster mergers on CCGs, the relaxation of the CCGs after merger events, and the implications of the observations for the size of the dark halos around the galaxies. Such questions will need to be answered if we are to discriminate between mergers or the processes of violent relaxation being primarily responsible for the behaviour we observe.

## ACKNOWLEDGMENTS

The authors thank Carolin Crawford, Harald Ebeling and Roderick Johnstone for discussions and comments on the manuscript. SWA acknowledges receipt of an S.E.R.C. Postdoctoral Fellowship. ACF thanks the Royal Society for support.